\documentclass[11pt]{article}

\usepackage[a4paper,margin=1in]{geometry}
\usepackage[utf8]{inputenc}
\usepackage[T1]{fontenc}
\usepackage{lmodern}
\usepackage{amsfonts}
\usepackage{amsmath}
\usepackage{amssymb}
\usepackage{amsthm}
\usepackage{booktabs}
\usepackage{enumitem}
\usepackage{graphicx}
\usepackage{hyperref}
\usepackage{microtype}
\usepackage{placeins}
\usepackage{nicefrac}
\usepackage{url}
\usepackage{xcolor}
\definecolor{safebg}{rgb}{0.85,0.93,0.85}
\definecolor{safebd}{rgb}{0.3,0.6,0.3}
\definecolor{unsafebg}{rgb}{0.95,0.85,0.85}
\definecolor{unsafebd}{rgb}{0.7,0.3,0.3}
\definecolor{pendingbg}{rgb}{0.92,0.92,0.95}
\definecolor{pendingbd}{rgb}{0.5,0.5,0.7}
\usepackage{tikz}
\usepackage{pgfplots}
\usetikzlibrary{positioning, arrows.meta, shapes.geometric, fit, calc, matrix}
\pgfplotsset{compat=1.18}
\usepackage[numbers]{natbib}

\newtheorem{definition}{Definition}
\newtheorem{theorem}{Theorem}
\newtheorem{proposition}{Proposition}
\newtheorem{corollary}{Corollary}
\newtheorem{observation}{Observation}

\title{Conflux: Preventing Privilege Escalation with Principal Contexts and System-Level Evaluation}

\author{Raya Buckley \\ University of Oxford \and Christian Schroeder de Witt \\ University of Oxford}

\date{}

\begin{document}

\maketitle

\begin{abstract}
LLM agents combine information from multiple principals while acting
externally. We formalise privilege escalation (PE) as the system-level security
objective: an action may proceed only if every principal that influenced it
would allow it. Model-level defences remain probabilistic, so enforcement must
sit outside model judgement. We present Conflux, a principal-context defence.
Its kernel, ITES, derives authority from provenance and the
organisation's existing access-control system and intersects the permissions of
every influencing principal. We prove a conditional authority-confinement
theorem: under complete mediation, no-laundering provenance, a known non-empty
decision context, fresh permission checks, and action binding, the kernel admits
no action that escalates privilege with respect to any recorded influencing
principal. The theorem assumes accurate provenance for every input that
influences the model.

Conflux extends the kernel with independent policy ports, argument-aware
authorisation, disclosure, consent, delegation, restriction, and exposure
planning. We describe two system-level evaluators. SLED is a black-box
evaluator: it exhaustively explores a declared finite model of a defence up to a
bound. SLED-V is a white-box evaluator: it checks a serialisable
intermediate representation against security and provenance properties. The
evidence is reported in the appendices: SLED and SLED-V results with the scope
of those runs; the AgentDojo runs available here are read-only and are not
effect-path evidence.
\end{abstract}

\section{Introduction}
\label{sec:intro}

Large language model (LLM) agents read documents, invoke tools, and perform
externally visible actions on behalf of users. They process natural-language
inputs from multiple principals \citep{earlycategorization}. A principal is a
user or other agent that can author information; documents, databases, and APIs
are sources that count through their recorded author.

Prompt injection is one mechanism by which untrusted information can change an
agent's behaviour \citep{struq,spotlighting,neuralexec}. Model-level defences
improve instruction following but remain probabilistic: adaptive attacks can
defeat empirical robustness
\citep{criticaleval,detectprompt,poisoningalignment,checkpointgcg}. A
system-level defence cannot generally prevent arbitrary prompt injection without
preventing the agent from processing untrusted information. We therefore treat
preventing \emph{privilege escalation} as the system-level security objective.
The relevant question is not whether an input is malicious, but whether every
principal influencing an action is authorised for that action. The enforcement
point is outside the model: an action is admitted only when every principal in
its recorded decision context is authorised for that action under the
organisation's current access-control system (ACS).

The central rule is simple. For an execution $e$ and an action $a$, let
$\mathrm{DC}(e,a)$ be the set of principals whose recorded information reaches
$a$ and its authority-bearing arguments. The defence kernel, which we call
\emph{ITES} (influence tracking with extrapolated security), permits $a$ only if
$\mathrm{DC}(e,a)$ is known and non-empty and every principal in
$\mathrm{DC}(e,a)$ satisfies the current permission relation $P(u,a)$. The
kernel does not consult the model. Model judgement may select a proposal, but it
cannot widen authority.

The Ava/Ben report workflow makes the distinction concrete. Ava asks an agent
to update a customer record with the remediation status described in a report
authored by Ben. The report supplies the status value, not the tool, recipient,
or target; refusing every data-dependent argument would reject a useful
workflow. After the report is processed, both Ava and Ben influence the action's
authority-bearing argument, so ITES permits it only when both are authorised
under the ACS. CaMeL instead derives control flow from the trusted user query and prevents
untrusted data from modifying it \citep{camel}; it enforces capability policies at
the tool boundary but does not state a PE guarantee. ITES admits arbitrary
data-dependent influence and constrains the authority of the resulting action.
The theorem assumes accurate provenance for every input to the model, including
inputs that select the tool, branch, recipient, or plan step.

Conflux is the complete defence: ITES plus the extensions that make the rule
usable in an agent architecture. It separates read, authorisation, visibility,
and consent decisions; adds argument-aware policy, disclosure, attribution,
delegation, restriction, and exposure planning; and keeps every extension
downstream of the kernel. Delegation is specified over finite models but is not
enabled at runtime, so the confinement theorem covers the static permission
relation.

This paper makes four contributions.
\begin{enumerate}
  \item \textbf{Formalisation.} We formalise the threat model around privilege
    escalation and give a common language for model-level and system-level
    defences and evaluators. The formalisation separates what a model can
    decide from what the system can enforce.
  \item \textbf{A system-level defence.} We present ITES, a provenance-based
    authority kernel, and Conflux, the architecture that extends it. We prove
    authority intersection, monotonicity, a context-relative maximality
    property, and a conditional confinement theorem.
  \item \textbf{A black-box system-level evaluator.} We present SLED, which
    evaluates a declared finite model of a defence independently of language
    model behaviour by exhaustively exploring the behaviours the model admits
    up to a bound.
  \item \textbf{A white-box system-level evaluator.} We present SLED-V, which
    re-expresses the transition system in a serialisable intermediate
    representation (IR) and checks security and provenance properties with
    bounded solver routes.
\end{enumerate}

The guarantee is safety-only. Conservative provenance and independent policy
dimensions can deny admissible tasks, so this paper does not claim utility or
liveness. The theorem is conditional on five assumptions stated in
Section~\ref{sec:ites}, including accurate provenance for every input that
influences the model.
The evidence is bounded and partial: finite-model results are statements about
declared models, and the AgentDojo runs exercise read-only paths rather than
effect-bearing defence. The AgentDojo runs are therefore a scope limit, not
effect-path evidence.

Section~\ref{sec:formal} gives the formal language and the threat model.
Section~\ref{sec:defences} presents related defences, the ITES kernel, the
Conflux extensions, and exposure planning. Section~\ref{sec:evaluators}
describes the related evaluators, SLED, and SLED-V. Section~\ref{sec:discussion}
discusses scope and work in progress; Section~\ref{sec:limitations} states the
limits; Section~\ref{sec:conclusion} concludes.

\section{Formalisation}
\label{sec:formal}

\subsection{Principals, executions, and tasks}
\label{sec:model}

Let $U$ be a finite set of principals and $D$ a set of data objects. An
\emph{action kind} $a \in A$ is an externally visible operation without argument
values; argument-level checks are a separate policy dimension. An \emph{input}
is any content tokenised for the model: a prompt, document, tool output, or
model message. The \emph{environment} is the tuple $(A,U,D,P,W,R,S)$, where
$P_t \subseteq U \times A$ is the permission relation at time $t$,
$W \subseteq D \times U$ records authorship, $R \subseteq D \times U$ records
read access \citep{abac}, and $S$ records the sources and provenance of data
objects. For principal $u$ and action $a$, $P_t(u,a)$ holds iff $u$ is
authorised to perform $a$ at $t$.

An \emph{execution} $e$ is a finite sequence of reads, model calls, and action
events. A read adds a data object and its recorded provenance to the execution.
A model call is a stateless function $M:\Sigma^* \to \Delta(A \times \Sigma^*)$
from the current token history to a distribution over proposals; the model may
be arbitrary and is never trusted. The token history includes all prior inputs,
so the model can condition on them, but the function itself has no state. An
action event $(e,a,t)$ dispatches action $a$ at time $t$. The mediator reads the
concrete action and arguments at the effect boundary, not the model's description
of them.

A \emph{plan} is a finite tree of reads, model calls, and actions. Planning
chooses the order and branching of reads and actions; it does not add authority.
A \emph{delegation} is a relation $(i,b,a,r)$ between an issuer $i$, a
beneficiary $b$, an action $a$, and an optional resource bound $r$; it is
effective only when recorded, unexpired, unrevoked, and within the issuer's
authority. The runtime result in this paper treats delegation as a finite-model
specification, not as an enabled permission source.

A principal is \emph{benign} or \emph{adversarial}. Benign principals
contribute to the task's goal actions; adversarial principals contribute to
attacker actions. A \emph{goal action} is an action required by the task, an
\emph{incidental goal action} is an action that is not required but contributes
to the goal, and an \emph{attacker action} is an action that serves an
adversary. An attacker action need not violate privilege escalation: an
adversary may use authority it already holds. The defence targets escalation,
not the existence of adversarial influence.

A \emph{task} $T$ is a set of acceptable executions. A task is \emph{secure}
under a defence if every execution the defence admits that completes $T$ has no
privilege-escalating action event. This is a safety property. It does not assert
that a secure task is completed, and it does not assert that every refusal is
correct.

\subsection{Threat model and security objective}
\label{sec:threat}

We consider an agent that performs externally visible actions on behalf of
users. The agent receives information directly through prompts and indirectly
through documents, APIs, and tool outputs. Any principal may provide arbitrary
inputs; we make no assumptions about the model's ability to distinguish
instructions from data or resist prompt injection. An adversary controls the
inputs of a subset $U_{\mathrm{adv}} \subseteq U$ but cannot alter the ACS, the
kernel, or the provenance store. The adversary's goal is an action event at
which some influencing principal is not authorised for the action. Collusion
between two individually authorised principals is a separate policy question.

\emph{Prompt injection} is an input that causes the model to treat data as an
instruction, or to let data select a tool, branch, recipient, or plan step. In
this model it is one mechanism by which an adversarial principal enters the
influence relation; the defence does not need to classify the input as an
injection. Delegation is outside the present runtime result.

Fix an execution $e$ with a dependence relation $\rightsquigarrow_e$ over data
items and action events. The relation is the least relation closed under the
execution's data-dependence and control-dependence rules; it is a parameter of
the model, not inferred from the action's value arguments alone. For an action
event $(e,a)$, define
\[
\mathrm{I}_{\mathrm{sem}}(e,a)
= \{u \in U \mid \exists d.\; u \in A(d) \text{ and } d \rightsquigarrow_e a\}.
\]
A principal semantically influences $a$ when an input carrying that principal's
recorded authority is a dependence predecessor of the action. The dependence
relation includes data and control dependence. The recorded set
$\mathrm{I}_{\mathrm{rec}}(e,a)$ is the runtime's approximation to this set. A2
states the accuracy condition
$\mathrm{I}_{\mathrm{sem}}(e,a) \subseteq \mathrm{DC}(e,a)$. It assumes that the
provenance store records every input that causally influences the model's
proposal or the action, including control inputs and any relay that controls
which messages reach the model. A control input that authors no data item is
represented by a control record whose author is the controlling principal. A2
assumes such records are present; it does not derive them.

Influence records a provenance contribution, not a permission. Authority is
normative: it records what the ACS permits. A role or model attribution may
describe why a principal participates, but it cannot remove an influencing
principal from the decision context or supply an approval.

\begin{definition}[Privilege escalation]
\label{def:escalation}
An action event $(e,a,t)$ \emph{escalates privilege} iff there exists an
influencing principal $u \in \mathrm{I}_{\mathrm{sem}}(e,a)$ such that
$\neg P_t(u,a)$, where $P_t$ is the effective access-control relation at
dispatch time. A trace escalates privilege iff at least one of its action
events does.
\end{definition}

Privilege escalation is action-level: it is a property of one executed action
event, and trace-level escalation is its existential lift. Under this
formulation, security depends only on the event's influence set and the current
ACS state, not on the source or mechanism of influence. Prompt injection is one
possible source; the defence treats it uniformly with any other.

\paragraph{Scenario taxonomy.}
Four scenarios recur in the rest of the paper.
\emph{Delegation.} Ava may perform $a_1$ or $a_2$; Ben has no permission.
Ava explicitly authorises Ben for $a_1$ and nothing else. Ben's authority
increases through a recorded grant, and the grant cannot exceed Ava's
authority. Delegation is secure when it is explicit, attenuated, and recorded.
\emph{Planning.} Ava may perform $a_1$ or $a_2$; Ben may perform many actions
but wants to launder one through the agent. Reading Ben's input before acting
can block Ava's action, while acting before reading Ben's input can complete
Ava's task without admitting Ben's action. Planning trades utility for
exposure; it does not widen authority.
\emph{Dynamic non-user input.} Ava authors a document and asks the agent to
follow it. The document is data, but Ava is its principal. If an attacker
cannot author the document, the task can be secure; if the attacker can, the
same text is unsafe. The distinction is authorship, not surface form.
\emph{Consent.} Ava consents to an action; Ben holds the permission the action
requires; the task needs both inputs. The agent may act on Ava's information,
and on Ava and Ben's information together, but not on Ben's alone. Consent is
an independent port, not a source of authority.
These scenarios are examples of the formal model, not additional objectives.

\subsection{A common language for defences and evaluators}
\label{sec:baseline}

We distinguish two kinds of information. \emph{Model-level information} is
information represented by, or produced by, the language model: token
probabilities, hidden states, an instruction/data classification, a refusal, or
a model-authored plan. \emph{System-level information} is information outside
the model: principal identity, provenance, ACS permissions, argument selectors,
consent records, and the identity of the action at the effect boundary.

A defence is \emph{model-level} when its admission decision depends only on
model-level information and the user request. A defence is \emph{system-level}
when its admission decision depends on system-level information at the effect
boundary. A \emph{mixed} defence uses both: a system-level predicate decides
whether an action is admissible, and a model-level rule selects among the
admissible actions.

\begin{proposition}[System-level necessity]
\label{prop:system-necessity}
Under worst-case model behaviour, no deterministic defence whose admission
decision depends only on model-level information and the user request can
guarantee that an unauthorised action is never admitted.
\end{proposition}
\begin{proof}
Take two executions with the same model-visible token history and the same
user request. In the first, a document is authored by principal $u$; in the
second, the same text is authored by an adversary $v$. Let the requested action
be authorised for $u$ but not for $v$. A deterministic model-level defence sees
the same information in both executions and must return the same decision. If
it admits the action, it admits an unauthorised action in the second execution;
if it refuses, it refuses an authorised action in the first. Thus no such rule
can guarantee confinement in both. The missing information is the recorded
authorship, which is system-level.
\end{proof}

\begin{proposition}[Model-level utility]
\label{prop:model-utility}
System-level information alone does not select among actions that are all
authorised. Two tasks can have the same principal context and the same ACS
state, while only one of two authorised actions completes the intended task.
Distinguishing them requires model-level information or a user-supplied goal.
\end{proposition}
\begin{proof}
Construct two tasks with identical principals, permissions, and available
actions. Let both tasks admit actions $a_1$ and $a_2$ under the ACS, but let
only $a_1$ complete the first task and only $a_2$ complete the second. A rule
that observes only the context and the ACS cannot choose between $a_1$ and
$a_2$; it must either refuse both or choose without task information. The same
argument applies when the two tasks differ only in whether the user's request
is genuine: the system state is identical, and the distinction is carried by
the request or by model-level interpretation.
\end{proof}

\begin{corollary}[Mixed-level decomposition]
\label{cor:mixed}
A mixed defence is the conjunction of a system-level admissibility predicate
and a model-level selection predicate. Neither pure component alone establishes
both confinement and task completion.
\end{corollary}
\begin{proof}
The system-level predicate can enforce confinement but cannot select a goal
action among admissible actions (Proposition~\ref{prop:model-utility}). The
model-level predicate can select a goal action but cannot enforce confinement
(Proposition~\ref{prop:system-necessity}). Their conjunction inherits both
properties when each component is sound for its part.
\end{proof}

The decomposition is not an equivalence. We do not claim that every mixed
defence can be represented as a pair of pure defences, or that a pure
system-level rule and a pure model-level rule together dominate every mixed
design. The open question is whether the two information classes are jointly
expressive enough under a fixed task model; we return to it in
Section~\ref{sec:in-progress}.

The two boundary cases make the division explicit. A \emph{refuse-all} defence
prevents privilege escalation but completes no task. A \emph{do-all} defence
completes every goal action but admits every attacker action that the ACS
permits. A usable defence must be neither: it must restrict authority without
discarding the task. The rest of the paper uses this language to compare
defences and evaluators.

\begin{table}[t]
\centering
\small
\begin{tabular}{@{}p{0.18\linewidth}p{0.34\linewidth}p{0.40\linewidth}@{}}
\toprule
Instrument & Question & What it cannot establish \\
\midrule
Model benchmark & Does a model resist a fixed attack suite? & Authority confinement under arbitrary model behaviour \\
System-level evaluator & Does a declared defence model satisfy its property for every admitted behaviour? & Implementation conformance; deployment safety \\
Policy audit & Does the ACS encode the intended restrictions? & Provenance completeness; model behaviour \\
\bottomrule
\end{tabular}
\caption{Three different questions. A model benchmark measures model behaviour;
a system-level evaluator measures a declared defence model; a policy audit
measures the ACS. No instrument substitutes for another.}
\label{tab:instrument-questions}
\end{table}

\section{Defences}
\label{sec:defences}

\subsection{Related defences}
\label{sec:related-defences}

Model-level defences change what the model sees, what it emits, or how its
output is filtered. StruQ uses structured queries to separate instructions from
data \citep{struq}; spotlighting marks untrusted spans \citep{spotlighting};
SecAlign and preference-based methods train the model to prefer safe behaviour
\citep{secalign}; DataSentinel and detection methods classify prompt-injection
attempts \citep{datasentinel,detectprompt}. These defences can reduce attack
success, but their decision is a function of model-level information. Under the
worst-case model of Section~\ref{sec:threat}, none can guarantee that an
unauthorised action is never admitted (Proposition~\ref{prop:system-necessity}).

System-level defences move the decision outside the model. The Dual LLM pattern
separates privileged planning from untrusted parsing
\citep{willison2023dual,designpatterns}. CaMeL adds capability tracking and
system-supplied policies enforced by a custom interpreter \citep{camel}. PFI
combines isolation, secure untrusted-data processing, and privilege guardrails
\citep{pfi}. Fides tracks confidentiality and integrity labels and enforces fixed
information-flow policies \citep{fides}; it does not state a PE guarantee. Wu et al.\ apply information-flow control across
the pipeline \citep{ifcdefense}. Progent mediates tool calls with symbolic
privilege control \citep{progent}; PACT records argument-level provenance
\citep{pact}; FORGE enforces history-sensitive Datalog policies \citep{forge};
PACE pairs path confinement with capability and effect verification
\citep{pace}; Agenra provides a policy language with a mechanised enforcement
core \citep{agenra}; and CaMeLoT adds pre-execution checking of temporal plan
properties to CaMeL \citep{camelot}. Several systems re-authorise at the effect
boundary: Zhu and Wang recheck protected effects at an atomic commit point
\citep{zhu2026runtime}, and APEX enforces authorised-effect and
task-endorsed-influence conditions from a contract compiled before untrusted
execution \citep{zheng2026apex}.

The defence kernel shares PACE's choice to keep the confinement decision outside
model judgement, but derives authority differently. An action's permission is
the intersection of the current permissions of every influencing principal in
the organisation's existing ACS, not a permission compiled from one request
contract, one task contract, or one application permission ledger. Because the
principal set is itself the label, ITES introduces no separate label or policy
language. Provenance accuracy and current ACS state remain assumptions.

The authority rule is a principal-sensitive form of Biba's low-water-mark
contamination rule: consuming information from an additional principal can
preserve or reduce effective authority but cannot increase it \citep{biba1977}.
LOMAC brought that rule to commodity systems and exposed the same
utility--security tension \citep{lomac}. ITES is structurally analogous but not
equivalent. Low-water-mark integrity lowers a subject's effective integrity
after it consumes lower-integrity information; ITES retains the identities of
all influencing principals and intersects their current ACS permissions. The
attenuation is induced by organisational authority, not by assigning one
integrity label to the execution. When permission sets are ordered by inclusion,
the intersection is their meet \citep{denning1976}; the principal-set authority
is therefore a writers-lattice construction, not a new lattice. A principal that
reads a value and passes it to the model is recorded in the author set of the
resulting value, so the label is the set of provenance writers, not the set of
readers.

The trusted-kernel separation also follows systems that made information-flow
labels explicit at the operating-system boundary. HiStar enforces labels in a
small kernel \citep{zeldovich2006histar}; Flume applies decentralised
information-flow control to standard OS abstractions \citep{krohn2007flume};
Asbestos provides kernel-enforced labels and event-process isolation for
multi-principal services \citep{efstathopoulos2005asbestos}. Clark--Wilson
supplies the complementary idea of certified transformations and separation of
duties \citep{clark1987comparison}, which frames the future question of when a
conservative influence set may be reduced without letting untrusted input choose
the transformation. Sabelfeld and Myers supply the vocabulary for explicit and
implicit flows \citep{sabelfeld2003}, and the decentralised label model supplies
multi-principal policies in which owners and readers constrain flows
\citep{myersliskov}. Declassification and endorsement
\citep{declassification-survey,robust-declassification,attacker-control,nonmalleable-ifc}
inform Conflux's visibility and disclosure design. Conflux applies this lineage
to agents with authenticated principal identities and authority taken from the
organisation's current ACS state.

\begin{table}[t]
\centering
\small
\setlength{\tabcolsep}{4pt}
\begin{tabular}{@{}p{0.16\linewidth}p{0.19\linewidth}p{0.19\linewidth}p{0.18\linewidth}p{0.18\linewidth}@{}}
\toprule
 & Model-level & Dual LLM & CaMeL & Conflux \\
\midrule
Decision locus & model input/output & planner/parser split & interpreter + policies & provenance + ACS \\
Information used & model-level & model + isolation & capabilities + policies & principal identities, provenance, ACS \\
Authority source & learned behaviour & planner policy & capability policy & existing ACS \\
Separate policy language & no & no & no (Python callables) & no \\
Tracks & none & data flow & capabilities & principals \\
Security objective & resist prompt injection & resist prompt injection & resist prompt injection and policy violation & prevent privilege escalation \\
\bottomrule
\end{tabular}
\caption{Conceptual comparison of model-level and system-level designs in the
common language of Section~\ref{sec:baseline}. The table is illustrative, not
exhaustive. ``Model + isolation'' means that the model still selects within an
isolated component. CaMeL policies are Python callables rather than a separate
policy language. Conflux's confinement decision does not depend on model
judgement; the model still selects a proposal.}
\label{tab:comparison}
\end{table}

\begin{table}[t]
\centering
\footnotesize
\setlength{\tabcolsep}{3pt}
\begin{tabular}{@{}p{0.15\linewidth}p{0.20\linewidth}p{0.20\linewidth}p{0.17\linewidth}p{0.20\linewidth}@{}}
\toprule
\textbf{System} & \textbf{Authority source} & \textbf{Tracked information} & \textbf{Decision point} & \textbf{Native property vs PE} \\
\midrule
Dual LLM & planner policy & data/control isolation & planner/parser split & injection resistance; no PE guarantee \\
CaMeL & developer capability policies & capabilities, data flow & interpreter at tool call & control-flow integrity and policy compliance; no PE guarantee \\
PFI & data-trust policy and user approval & data references & guardrail before privileged call & PE objective; unsafe-flow detection, no PE guarantee \\
Fides & per-tool label policies & confidentiality and integrity labels & planner/tool wrapper & integrity noninterference and explicit secrecy; no PE guarantee \\
Progent & symbolic privilege policy & tool and argument privileges & tool call & privilege policy; no PE guarantee \\
PACT & contracts and policies & argument provenance & tool call & argument policy; no PE guarantee \\
FORGE & Datalog policies & execution history & tool call & history policy; no PE guarantee \\
Conflux & existing ACS & principal provenance & effect boundary & PE under accurate provenance \\
\bottomrule
\end{tabular}
\caption{System-level designs by the information they track and the point at
which they decide. The final column separates a defence's native property from
the property evaluated here. ``No PE guarantee'' means the mechanism can prevent
some escalation but does not state the principal-sensitive property: every
influencing principal's current ACS permission is checked for the action. The list
is representative, not exhaustive;
Section~\ref{sec:related-defences} discusses the mechanisms and their limits.}
\label{tab:system-comparison}
\end{table}

\subsection{Defence kernel (ITES)}
\label{sec:ites}

The kernel rests on one principle: every execution inherits the effective
authority implied by the provenance of its inputs. For an execution $e$ whose
direct inputs are data objects $ds \subseteq D$, the \emph{Principal Context} is
\[
\mathrm{PC}(e) = \bigsqcup_{d \in ds} A(d),
\qquad
A(d) =
\begin{cases}
\{u \in U \mid W(d,u)\}, & \text{if $d$ has a recorded author},\\
\{\bot\}, & \text{otherwise},
\end{cases}
\]
where $\bigsqcup$ merges author sets and propagates $\bot$: if any input has
unknown provenance, the whole context is unknown. An object produced by an
execution inherits its complete Principal Context.

Authorship alone is not enough. For a data object $d$, let
$\mathrm{Src}(d) \subseteq D$ be the objects recorded as its direct sources. The
provenance store is \emph{no-laundering} when, for every $d$,
\[
\bigcup_{d' \in \mathrm{Src}(d)} A(d') \;\subseteq\; A(d).
\]
This condition is an assumption on the store, not a consequence of the
authorisation rule. Without it, trusted Ava can copy attacker-authored text
into a new document, the new document records only Ava as author, and the
attacker disappears from the context. An action derived from that document can
then be permitted although the attacker would not be. Only under this condition
is recorded provenance a conservative approximation of influence.

Given an execution $e$ and an action $a \in A$, let the \emph{decision context}
be the merge of the Principal Context with the provenance of the action's
authority-bearing arguments:
\[
\mathrm{DC}(e,a) = \mathrm{PC}(e) \cup \mathrm{Prov}_{\mathrm{arg}}(a),
\]
where $\mathrm{Prov}_{\mathrm{arg}}(a)$ is the set of principals recorded for
the action's authority-bearing arguments. The kernel, ITES, permits $a$ only if
$\mathrm{DC}(e,a)$ is known and non-empty and
\[
\forall u \in \mathrm{DC}(e,a),\; P(u,a),
\]
and the independent read, visibility, consent, and argument-policy decisions
also allow. Equivalently, the effective authority is the intersection of the
permission sets of all principals in the decision context. The rule is stated
only for a decision context that is non-empty and known (A5); an execution whose
context contains no principal is not a valid input to the decision. The read,
visibility, consent, and argument-policy dimensions are independent decisions
(Section~\ref{sec:conflux}) and can only deny further. The rule applies to
externally effectful actions; Section~\ref{sec:scope} delimits the action
classes outside its conclusion.

\begin{observation}[Permit implies pointwise permission]
\label{lem:permit-pointwise}
For a known, non-empty decision context $X$ and an action $a$, if the
authorisation dimension of the kernel permits $a$ on $X$, then $P(u,a)$ holds
for every $u \in X$.
\end{observation}
This is a restatement of the rule, not an independent property of an
implementation; we record it only to make the two uses below explicit.

\begin{proposition}[Monotonicity]
\label{prop:monotonicity}
For a known, non-empty principal set $X \subseteq U$, write
\[
\mathrm{Permitted}_P(X) = \{a \in A \mid \forall u \in X,\; P(u,a)\}.
\]
Adding a principal can only add constraints:
\[
X \subseteq Y \quad\Longrightarrow\quad
\mathrm{Permitted}_P(Y) \subseteq \mathrm{Permitted}_P(X).
\]
\end{proposition}
\begin{proof}
Take $a \in \mathrm{Permitted}_P(Y)$. Then $P(u,a)$ holds for every $u \in Y$.
Since $X \subseteq Y$, it holds for every $u \in X$, so
$a \in \mathrm{Permitted}_P(X)$. Applied to decision contexts, if
$\mathrm{DC}(e_1,a) \subseteq \mathrm{DC}(e_2,a)$, then
$\mathrm{Permitted}_P(\mathrm{DC}(e_2,a)) \subseteq
\mathrm{Permitted}_P(\mathrm{DC}(e_1,a))$: influence accumulated by an
execution can preserve or reduce authority, never increase it.
\end{proof}

\begin{proposition}[Context-relative maximality]
\label{prop:maximality}
Among sound rules that use only the recorded decision context and the ACS, the
authorisation dimension of ITES is maximal: it admits exactly the actions that
every principal in $\mathrm{DC}(e,a)$ permits.
\end{proposition}
\begin{proof}
Soundness follows from Observation~\ref{lem:permit-pointwise}. For maximality, let
$R$ be another rule that uses only the same context $X$ and the same relation
$P$. If $R$ admits an action $a \notin \mathrm{Permitted}_P(X)$, then some
$u \in X$ lacks $P(u,a)$, so $R$ is unsound. If $R$ denies an action
$a \in \mathrm{Permitted}_P(X)$, then $R$ is strictly more restrictive than
ITES on the same information. Hence no sound rule in this class admits more
than ITES. The statement is relative to the recorded context and to the
authorisation dimension only: the full kernel decision is smaller when an
independent read, visibility, consent, or argument-policy port denies, and a
more precise context can admit actions that ITES denies.
\end{proof}

\paragraph{Authority confinement.}
The following theorem is a soundness property of the rule: any runtime that
satisfies A1, A2, A4, A5 and conforms to the rule confines the authority of
every effectful action it mediates. A3 is a conformance assumption, not a
derived property; the theorem does not establish that a particular
implementation satisfies it. That gap is a refinement obligation, stated in
Section~\ref{sec:scope}.

\begin{theorem}[Authority confinement]
\label{thm:confinement}
Let $e$ be an abstract execution, let $t$ be a dispatch time, and let $a$ be an
effectful action in the domain of the intersection rule. Assume:
\begin{enumerate}
  \item[(A1)] \emph{Mediation.} Every action in the domain of the rule is
        mediated by ITES, and executes only if the kernel returns \emph{permit}
        for $\mathrm{DC}(e,a)$ on the authorisation dimension and the
        independent read, visibility, consent, and argument-policy decisions
        also permit;
  \item[(A2)] \emph{Accurate provenance.} The provenance store records every
        input that causally influences the model's proposal or the action,
        including control inputs that select the tool, branch, recipient, or
        plan step, and any relay that controls which messages reach the model.
        For each such input, the record identifies the principal that supplied
        it. The recorded influence set therefore satisfies
        $\mathrm{I}_{\mathrm{sem}}(e,a) \subseteq \mathrm{DC}(e,a)$, and the
        provenance store is no-laundering in the sense above;
  \item[(A3)] \emph{Rule conformance.} The runtime's authorisation decision for
        an action is the ITES intersection rule on $\mathrm{DC}(e,a)$;
  \item[(A4)] \emph{Action binding and recheck.} The action executes only after
        the mediation service has re-derived the decision for
        $\mathrm{DC}(e,a)$ and checked that the presented certificate matches
        the authorised decision, binding the action, the decision context, the
        executing branch, and the current policy versions. The recheck and
        dispatch are atomic with respect to policy changes, or dispatch fails
        closed if the policy version changes between them;
  \item[(A5)] \emph{Known, non-empty context.} $\mathrm{DC}(e,a)$ is known and
        non-empty, where $\mathrm{DC}(e,a) \subseteq U \cup \{\bot\}$, $\bot$
        is absorbing, and \emph{known} means $\bot \notin \mathrm{DC}(e,a)$.
\end{enumerate}
Then $e$ does not perform privilege escalation at action event $(e,a,t)$.
\end{theorem}
\begin{proof}
We prove the invariant by induction over the events of $e$. At each prefix,
let $C_a$ be the decision context rechecked for an action event $a$. The
invariant has two parts: (i) every principal that has contributed to the prefix
belongs to the current Principal Context; and (ii) every action event that has
executed had
$\mathrm{I}_{\mathrm{sem}}(e,a)\subseteq C_a$ and $P_t(u,a)$ for every
$u\in C_a$ at its dispatch time.

The base case is the initial input context. A read adds the author set $A(d)$
of the read object to the Principal Context. The no-laundering clause of A2
ensures that $A(d)$ contains the authors of every direct source of $d$, so the
read cannot remove a principal that already contributed; a model call produces
a proposal and adds no authority. Thus every non-action event preserves (i).

At an action event, A1 requires the kernel and every independent port to
permit. A3 makes the kernel's authorisation decision the intersection rule, so
Observation~\ref{lem:permit-pointwise} gives $P_t(u,a)$ for every
$u\in\mathrm{DC}(e,a)$. A2 gives
$\mathrm{I}_{\mathrm{sem}}(e,a)\subseteq\mathrm{DC}(e,a)$. A4 ensures that
the executed effect is the action whose context was checked and that the
recheck is atomic with the policy version, or fails closed; A5 ensures that the
context is known and non-empty. Hence every semantic influencing principal
satisfies $P_t(u,a)$ at dispatch time. By Definition~\ref{def:escalation}, the
event does not escalate privilege. Induction over $e$ gives the result for
every action event in $e$.

The argument is a soundness proof for the rule. It does not show that an
implementation satisfies A2 or A3; those are refinement obligations.
\end{proof}

The result is independent of the source of influence because A2 records every
causal input. Prompt injection, retrieved documents, and untrusted tool outputs
are treated uniformly: once information contributes to an execution, the
authority of that execution is constrained by the permissions of the
corresponding principals.

\subsection{Scope of the guarantee}
\label{sec:scope}

The theorem is a rule-level conditional result: it does not establish that a
given implementation conforms to the rule's operational semantics. The proof
depends on the provenance relation, the decision context $\mathrm{DC}(e,a)$, and
the execution-time permission relation $P_t$, not on a particular transition
relation. An \emph{effectful action} is one that performs an external effect,
including the effects of a nested execution. A message is a disclosure event
governed by visibility and consent, not by per-principal $P$; messages lie
outside the theorem's conclusion, and covering them requires a per-principal
message permission in the decision context.

It also does not establish policy adequacy: whether the ACS encodes the
organisation's intended restrictions is a separate assurance question. It is not
a noninterference or confidentiality result. Side channels, implicit flows, and
read-based disclosure are outside A1--A5; read, visibility, consent, and
argument policy are independent decisions. Authentication is part of the trusted
computing base: A2 assumes accurate provenance for every input that reaches the
model and every influence on the action.

The result is safety-only. Conservative provenance and independent policy
dimensions can deny actions that the ACS would permit, so the theorem
establishes neither utility nor liveness. Permission is action-level: it does
not imply safe recipients, amounts, destinations, or argument values, and the
theorem is not an exfiltration result. Trust is action-relative: a principal may
be trusted for one action and untrusted for another, so the rule checks the
action at the effect boundary rather than assigning a global trust label. Delegation is also outside the result:
when delegation is enabled, $P$ must be the effective relation
$\mathrm{ACS}_{\mathrm{effective}}(e)$, not the explicit ACS.

A4 closes the certificate-binding gap at recheck. A3 is a separate
conformance obligation: the implementation must make the same authorisation
decision as the intersection rule. Certificates are unkeyed records: they detect
stale decisions and actions or arguments mutated after authorisation, but not a
compromised executor or a process able to rewrite both the certificate and the
action. A4 therefore requires recheck and dispatch to be atomic, or dispatch to
fail closed when the policy version changes between them; without that atomicity
a race between recheck and the external effect remains outside the guarantee.

\begin{table}[t]
\centering
\small
\begin{tabular}{@{}p{0.25\linewidth}p{0.31\linewidth}p{0.36\linewidth}@{}}
\toprule
Step & Benign request & Injected request \\
\midrule
Input provenance & Ava's request & Ava's request plus an external document \\
Principal Context & \{Ava\} & \{Ava, external\} \\
Policy result & Ava may create the event & Ava may create the event; external may not \\
Outcome & certificate, effect & denial, no effect \\
\bottomrule
\end{tabular}
\caption{A worked example of the authority rule. The injected request needs no
injection detector: the external principal enters the decision context, and the
action is denied because that principal lacks authority.}
\label{tab:conflux-example}
\end{table}

\subsection{Defence extensions (Conflux)}
\label{sec:conflux}

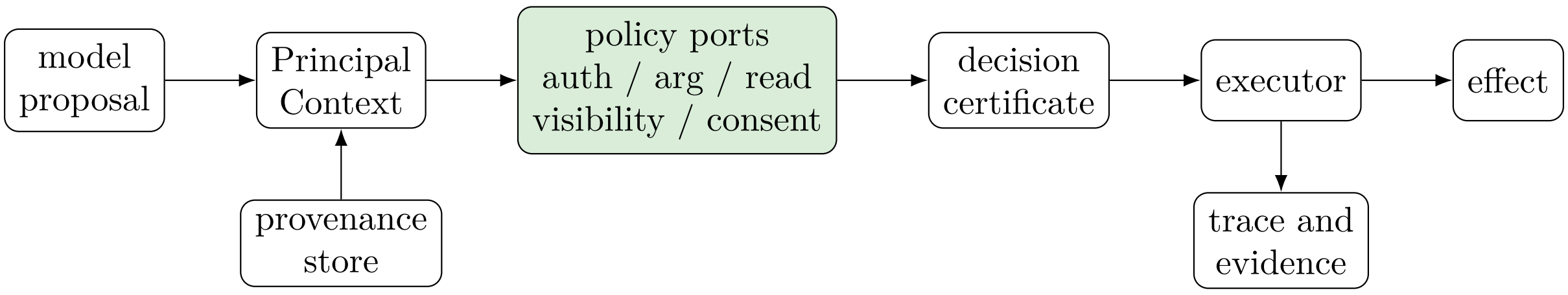
\begin{figure}[t]
\centering
\resizebox{\linewidth}{!}{%
\begin{tikzpicture}[node distance=7mm and 9mm, every node/.style={font=\small}, box/.style={draw,rounded corners,align=center,inner sep=4pt,minimum height=8mm}, kernel/.style={draw,rounded corners,fill=safebg,align=center,inner sep=4pt,minimum height=8mm}]
\node[box] (model) {model\\proposal};
\node[box,right=of model] (ctx) {Principal\\Context};
\node[kernel,right=of ctx] (ports) {policy ports\\auth / arg / read\\visibility / consent};
\node[box,right=of ports] (cert) {decision\\certificate};
\node[box,right=of cert] (exec) {executor};
\node[box,right=of exec] (effect) {effect};
\draw[-{Latex}] (model) -- (ctx);
\draw[-{Latex}] (ctx) -- (ports);
\draw[-{Latex}] (ports) -- (cert);
\draw[-{Latex}] (cert) -- (exec);
\draw[-{Latex}] (exec) -- (effect);
\node[box,below=of ctx] (prov) {provenance\\store};
\node[box,below=of exec] (audit) {trace and\\evidence};
\draw[-{Latex}] (prov) -- (ctx);
\draw[-{Latex}] (exec) -- (audit);
\end{tikzpicture}}
\caption{Conflux request path. ITES is the authority kernel; the model and
planner are untrusted proposal sources. The application mediation service
rechecks the certificate and current policy immediately before dispatch; the
executor is an effect sink, not an independent certificate verifier. The figure
is a design schematic, not a claim that every extension is runtime-enabled.}
\label{fig:conflux-architecture}
\end{figure}

Conflux is not a second defence. It combines ITES, the finite-model evaluators,
and a set of extensions with one dependency direction and one mediation
boundary: extensions depend on the kernel, never the reverse.
Figure~\ref{fig:conflux-architecture} shows the request path: model proposal,
Principal Context, independent policy ports, decision certificate, and
executor. The model and the planner remain outside the authority decision.

\paragraph{Fail-closed design and trust boundary.}
Conflux is fail-closed. Missing authority, an unknown principal, a stale
certificate, or an unsupported backend is denied or reported as
\texttt{UNKNOWN}, never silently allowed. Domain values are immutable, so a
decision cannot change after it is issued. Alternative branches share an
immutable parent and cannot borrow a sibling's certificate. The security kernel
is pure: it does not call the model, the network, or a provider. Because the
kernel is pure, SLED and SLED-V can explore the same transition relation without
semantic drift. A provider adapter is outside the kernel and is not a production
isolation boundary.

\paragraph{Request lifecycle.}
A read creates or extends provenance. A model call produces a proposal batch.
The application derives the Principal Context as the join of every principal
whose information reached the proposal. Each independent policy port then
answers its own question: whether the principal is authorised for the action;
whether each authority-bearing argument selector is authorised; whether the
principal may observe the input; whether the action is visible to the intended
audience; and whether there is consent for the disclosure or effect. All must
allow. A single denial blocks the action and names the rule that denied it.

A successful decision yields a certificate binding the action fingerprint, the
context fingerprint, the branch identity, and the policy versions. Execution
rechecks the certificate against the current policy. A certificate from another
branch, a changed policy, or a reused one-use grant fails closed. The outcome is
recorded as an event, and a refused proposal is still a security success: it
produced no effect.

\paragraph{Independent policy ports.}
Authorisation, read access, visibility, and consent are separate decisions.
Consent cannot override authority; visibility cannot override read policy. Each
port answers only its own question, and the effective decision is their
conjunction, so one dimension cannot manufacture authority for another. The
present kernel evaluates these decisions pointwise and per action. A general
interface would also admit a known, non-empty decision context, an exact action
fingerprint, per-branch trace state, and policy versions, and return allow,
deny, or pending with a replayable justification; that interface is future work.

\paragraph{Argument-aware authorisation.}
Trusted action-argument roles carry provenance independently of the proposing
execution. Selector policy checks each principal against each authority-bearing
argument component, so authority cannot be borrowed through content or consent.
The check is only as complete as the selector model: an argument that a tool
schema does not mark as authority-bearing is outside it, and widening the
selector model is a refinement obligation.

\paragraph{Disclosure, attribution, and consent.}
Deterministic redaction produces audience-specific event views without copying
hidden payload fields. Attribution is derived from provenance, context, and
policy records rather than from model statements, so a model explanation cannot
establish who caused an action. Consent is an independent port, so a consent
record cannot widen authority. These mechanisms constrain disclosure decisions;
they do not establish noninterference.

\paragraph{Scoped delegation.}
The delegation model binds issuer, beneficiary, operation, resource, and
argument, with expiry, revocation, and atomic one-use consumption. Delegation is
specified and verified over finite models but is not enabled at runtime: ITES
denies delegation operations, so the effective permission relation is the
explicit ACS. The confinement result therefore covers static permission only. Endorsement is
not modelled as a separate permission source here: it can change the recorded
author set, but it cannot widen the ACS.

\paragraph{Restriction.}
Restriction is the CaMeL/Dual LLM-style planning discipline: a privileged
planner may select only from an operation alphabet declared for the task, and
untrusted data cannot add operations to that alphabet. The executor receives
only operations that the kernel admits. Restriction is not a separate source of
authority; it reduces the proposal space before ITES decides. We model the
alphabet and the plan shape in the plan IR, but the runtime planner is a
research prototype and the finite-model results are statements about the
declared IR, not about an operational planner.

\subsection{Exposure planning}
\label{sec:exposure-planning}

Reading an additional principal can only shrink the set of permitted actions
(monotonicity). The order of reads therefore changes what an agent can do. An
agent that reads an untrusted document before a sensitive action may block that
action; the same agent may read the document after the action, or branch the
document into an isolated sub-execution, and complete the task without
escalation. Exposure planning makes this choice explicit: it selects the order
and branching of reads and actions so that every action is permitted by the
context that exists at that point. Treating read order and branching as part of
the defended plan is a design contribution; it does not change the authority
rule.

A planning domain is a finite tuple $(S,s_0,G,\mathcal{O},c)$. A state records
the principals acquired so far, the actions performed, and the current branch.
The operators are $\mathrm{read}(d)$, which adds $A(d)$ to the context, and
$\mathrm{act}(a)$, which is available only if every principal in the current
decision context satisfies $P(u,a)$. The goal $G$ is a set of acceptable final
states. The cost $c$ counts new principals, sensitive reads, irreversible
effects, and the number of branches. A plan is a finite tree of operators. It is
\emph{admissible} when every action along the plan is permitted at the point it
occurs. An \emph{optimal} admissible plan reaches $G$ with minimum cost.

Because the planning domain and horizon are finite, the set of bounded plans is
finite. If at least one admissible plan reaches $G$, the admissible plans that
reach $G$ form a non-empty finite set, so a minimum of the real-valued cost
exists; a deterministic tie-break selects one. The minimum need not be unique:
two plans can acquire the same principals and perform the same actions in
different orders at equal cost. We therefore require a declared tie-break
whenever the planner returns a single plan.

\paragraph{Hierarchical example.}
Suppose principal $A$ may read $d_A$ and perform $a_1$, principal $B$ may read
$d_B$ and perform $a_2$, and neither may perform the other's action. The task
requires $a_1$ using $d_A$ and $a_2$ using $d_B$. A plan that reads both
documents before either action has context $\{A,B\}$ and is refused for both
actions. A plan that performs $a_1$ after reading $d_A$, then reads $d_B$ and
performs $a_2$, is admissible if the actions do not need the other principal's
data. If the task instead requires one action that combines both documents, no
admissible plan exists. The example shows that planning can preserve utility
without widening authority: each action is checked against the context that
exists when it is performed.

Exposure planning does not enter the confinement theorem. It selects among
actions that ITES would already admit; every selected action is re-mediated at
dispatch. The finite-model results are in Appendix~\ref{app:sledv}; they cover
the declared plan IR, not an operational LLM planner.

\section{Evaluators}
\label{sec:evaluators}

\subsection{Related evaluators}
\label{sec:related-evaluators}

AgentDojo supplies prompt-injection tasks and metrics in realistic agent
protocols \citep{agentdojo}. It measures whether an agent completes user tasks
and whether an attack changes behaviour. It does not, by itself, establish that
a defence confines authority: a run can complete no effectful task and still
report no violation because no effect was attempted. The AgentDojo runs available to this
paper exercise read-only paths; they cannot test effect-path defence.

InjecAgent supplies a second injection channel: an attacker can offer a tool
alongside the user's tools \citep{injecagent}. It separates data stealing from
direct harm and classifies benchmark cases. It measures attack outcomes, not
provenance completeness or policy correctness.

Tau2-Bench evaluates conversational agents in a dual-control environment, where
the agent and the user both act on a shared tool state \citep{tau2}. It measures
task completion under user interaction and tool use. It does not observe
principal provenance or ACS permissions, so it cannot establish authority
confinement. It is useful for the capability side of the mixed-level
decomposition (Corollary~\ref{cor:mixed}), not for the system-level
predicate.

\begin{table}[t]
\centering
\small
\begin{tabular}{@{}p{0.16\linewidth}p{0.36\linewidth}p{0.40\linewidth}@{}}
\toprule
Instrument & Question & What it cannot establish \\
\midrule
AgentDojo & Does an agent preserve utility and stop targeted injections in realistic protocols? & Effect-path defence when the runs are read-only \\
InjecAgent & Which principal authorised an effect when an attacker offers a tool? & Live attack success or utility; it classifies cases only \\
Tau2-Bench & Can an agent complete dual-control tasks with a user and tools? & Provenance completeness; authority confinement \\
\bottomrule
\end{tabular}
\caption{The question each external evaluator answers. No evaluator
substitutes for another, and a benchmark result is not a system-level
confinement proof.}
\label{tab:benchmark-boundaries}
\end{table}

\subsection{Black-box evaluator (SLED)}
\label{sec:sled}

SLED is our black-box system-level evaluator. It does not execute a language
model. It takes a declared finite model of a defence: a set of principals, an
ACS state, data objects, tools, a transition relation, an initial state, and a
property. It then explores every behaviour the model admits up to a declared
bound.

The model treats the language model as an adversarial, nondeterministic
proposal source. At each model step, the evaluator may choose any well-typed
proposal that the declared interface admits; it never asks whether a particular
model would produce that proposal. An execution reads objects, produces derived
objects, propagates provenance, schedules nested executions, and may dispatch
effects. A transition applies the declared defence to the proposal. The search
is breadth-first and stops when the finite state space is exhausted, a
counterexample is found, the declared bound is reached, or the model cannot be
evaluated.

SLED reports four outcomes: the finite model is exhausted with no violation; no
violation is found within the declared bound; a counterexample is found; or the
model cannot be evaluated. A counterexample is a concrete action sequence that
reaches a violating state. A bounded outcome speaks only about the declared
model, not about an unbounded deployment.

Each explored trace is classified by observable system-level outcomes. Security
covers privilege escalation, unauthorised effect, and forbidden read; utility
covers task completion under the declared authority. The classifier uses the
declared environment, not hidden model state. This is why a model benchmark and
a system-level evaluator answer different questions
(Table~\ref{tab:benchmark-boundaries}). A benchmark measures a model on a fixed
suite; SLED measures a defence model over all behaviours the model admits up to
a bound.

For a declared finite model, trace enumeration and task-level enumeration agree
on reachable goal states only under a projection condition. We require that
(i) every trace projects to a task event sequence, (ii) every admitted task
event sequence lifts to at least one trace, and (iii) projection preserves goal
and action labels. Under (i)--(iii), every reachable goal state has a trace
whose projection reaches it, and every admitted task sequence that reaches a
goal state lifts to a trace that reaches it. This is a modelling condition, not
an empirical claim. If lifting fails, trace enumeration can miss tasks; if
projection fails, task enumeration can include tasks with no trace.

\begin{proposition}[Finite black-box evaluation]
\label{prop:blackbox-future}
A black-box evaluator that observes a defence only on finitely many interactions
cannot establish an unbounded property of that defence. An unbounded verdict
requires a declared model of the defence's future code.
\end{proposition}
\begin{proof}
Take two defences that agree on every interaction up to the declared bound and
differ after it. The evaluator sees the same finite behaviour and must return
the same verdict, so it cannot distinguish the safe from the unsafe defence. A
declared model breaks the tie only by fixing the future behaviour; the verdict
is then conditional on that model.
\end{proof}

Appendix~\ref{app:sled} reports the three newest archived SLED runs: the
future-code ablation, the final-depth batch sweep, and the environment-complexity
series. A separate provenance-precision fixture is reported in the same appendix
under its own heading; it is not a SLED state-space run. We do not reproduce the
older archived trace-count experiment here; its enumeration differs from the
current evaluator, so a comparison would not be like-for-like. Every result is a
statement about its declared model or fixture, not an unbounded security proof.

\subsection{White-box evaluator (SLED-V)}
\label{sec:fv}

SLED-V evaluates transition systems expressed in a serialisable intermediate
representation (IR) rather than arbitrary programs. The IR has a finite-domain
fragment, which supports depth-unbounded model checking, and a plan fragment,
which is checked at explicit node bounds. Because the IR is data, it admits
solver-backed routes. The IR has a reference
interpreter, so a solver verdict can be checked against a direct semantics
rather than trusted on its own. The properties cover unauthorised effects and
reads, provenance monotonicity, absence of implicit delegation, and bounded
resource use. A bounded result is a statement about a declared model at a
declared bound. An unbounded claim additionally requires a model checker over a
finite domain and an argument that the implementation conforms to the IR.

\paragraph{Solver routes on plan IRs.}
Two solver routes handle bounded plan obligations: predicate abstraction with
CEGAR refinement, and SMT-based induction. The plan IR declares no step range;
each plan fixes a node bound in its guards. A verdict is therefore a statement
at that bound, not a parametric proof for arbitrary plan depth. IC3/PDR requires
a finite domain; it rejects the plan IR's unbounded integer step counter, so it
applies to finite-domain models and is not a third plan-IR route.

A route returns \texttt{safe}, \texttt{unsafe}, or \texttt{unknown}. An
\texttt{unknown} is never reported as safety and does not extend to obligations
the reference leaves undecided. A positive verdict may
need an auxiliary invariant and a transition guard, because the property alone
is not inductive. Two routes can also disagree on the same obligation; a
bounded non-result then establishes nothing in either direction.

\paragraph{Falsifiability of a holding verdict.}
A holding verdict supports a security claim only if the invariant is
\emph{falsifiable} under a declared mutation alphabet: if weakening a single
rule guard cannot falsify the property, the check shows that the property is
entailed by the encoding, not that the defence enforces it. We classify each
invariant as \texttt{checked}, when it holds at baseline and some single-rule
guard mutation falsifies it at a reachable state;
\texttt{checked\_by\_construction}, when it holds at baseline and no such
mutation falsifies it; or \texttt{violated\_at\_baseline}, when it does not
hold at baseline. A \texttt{checked} class is a statement about the declared
mutation alphabet, not an absolute property of the invariant.

The classification is relative to an analysis as well as to the mutation
alphabet. When two analyses disagree about the same invariant, only the weaker
class supports a claim, so a \texttt{checked\_by\_construction} verdict does
not support a holding claim. A \texttt{checked} class can also be vacuous for
the rule it is attributed to: a guard mutation may falsify the invariant even
when no baseline transition satisfies the rule's original guard. The class then
records sensitivity to the mutation alphabet, not a demonstration that the
defence enforces the property.

The reference model is a one-principal, one-action Boolean instance: one
influencing principal, one action, one permission bit, with no general
principal set, action set, policy language, or provenance graph, and no
conformance argument from the implementation. A \texttt{checked} verdict there
supports the rule's block/allow partition in that instance, not the general
theorem.

Appendix~\ref{app:sledv} reports semantic conformance over bounded IR variants
and the archived IR property tables. The conformance check shows that the
reference interpreter and the optional solver backends agree on the tested
encodings; it does not show sensitivity to defect interactions.

\section{Discussion}
\label{sec:discussion}

\subsection{Scope of the authority claim}
The authority-confinement result is conditional on A1--A5. A2, accurate
provenance, is the most contestable assumption; A1 and A4 remain assumptions on
the trusted computing base; A3 is a conformance obligation on the
implementation. Section~\ref{sec:limitations} gives a concrete case in which
provenance is inaccurate and A2 fails. Action binding is a separate refinement
obligation: the mediator
revalidates a certificate, but the executor and the certificate channel remain
inside the trusted computing base.

The theorem is also narrower than a general access-control result. It applies
to executions that satisfy A2, over a known, non-empty decision context. The reference IR is a one-principal, one-action Boolean instance: the
rule's block/allow partition holds there, but this is not a proof about
arbitrary principal sets, policy languages, or provenance graphs. Authorised-read
safety is not noninterference: a system can obey local ACL checks and still leak
information through outputs, timing, or control flow
(Section~\ref{sec:limitations}).

The evidence matches this scope. Finite-model results are statements about
declared models, not deployment guarantees, and the benchmark runs so far
exercise read-only paths, so they establish neither defence efficacy nor
utility cost.

\subsection{Relation to prior work}
The authority-intersection rule is a meet in the powerset lattice and is
structurally analogous to low-water-mark integrity
\citep{biba1977,denning1976}. It is not a new lattice construction.
History-based access control derives authority from the execution that led to a
request \citep{abadi2003access}, and the confused-deputy problem motivates
capability-style checks that bind authority to the request's provenance
\citep{hardy1988confused}. Information-flow labels, argument-level provenance,
and effect-time re-authorisation are prior or concurrent work
\citep{camel,fides,pact,forge,zhu2026runtime,zheng2026apex}. PFI also frames
privilege escalation as the agent security objective \citep{pfi}. Our formulation
differs in the authority source and the quantified principal set: authority is the
intersection of the current ACS permissions of every provenance-recorded
influencing principal, and model judgement stays outside the authority trusted
computing base so that confinement does not depend on it. The information-flow
lineage supplies the vocabulary and the limits
\citep{sabelfeld2003,myersliskov,declassification-survey,robust-declassification,attacker-control,nonmalleable-ifc}.

CaMeL and Fides enforce control-flow integrity and information-flow policies at
tool boundaries, but neither states a PE guarantee: their policies are
per-tool/capability or label policies, not the intersection of every influencing
principal's current ACS permissions. LLMbda \citep{llmbda2026} gives a probabilistic noninterference
result for a provenance-tracked calculus; it does not define authority from an
external ACS. PRUDENTIA \citep{prudentia2026} optimises plans for policy compliance and autonomy, not
principal-set authority. Multi-CaMeL \citep{multicamel2026} preserves provenance
across agent boundaries, but it is a communication protocol, not an
authority-confinement result for an existing ACS.

\subsection{In progress}
\label{sec:in-progress}
Several extensions are in progress rather than established results.
\begin{itemize}
  \item \emph{Industry policy languages.} Integrating ITES with deployed policy
    languages would test whether the authority rule can use existing policy
    state without a translation layer.
  \item \emph{Larger AgentDojo runs.} The archived runs use small models and
    read-only paths. Effect-bearing runs with larger models are needed before
    the benchmark can support a defence or utility-cost claim.
  \item \emph{LLM planning and defence adoption.} It is open whether models can
    plan under the authority rule, use delegation and restriction well, and
    recover from refusals without losing the task.
  \item \emph{Provenance for control inputs.} Recording tool, branch, recipient,
    and plan-step selection in the provenance store is an implementation
    obligation under A2.
  \item \emph{Exact action binding.} Binding the authorised action to the
    executed action at the effect sink, including argument values and schema
    defaults, remains outside the current runtime guarantee.
  \item \emph{Policy synthesis and adequacy.} The authority rule does not tell
    an operator whether the resulting policy leaves enough capability for the
    task. Adequacy requires a task specification, an over-blocking budget, and a
    synthesis procedure.
  \item \emph{Noninterference and read confidentiality.} Authorised-read safety
    is not noninterference. Separating read authorisation from information-flow
    confinement, including covert channels, is future work.
  \item \emph{Multi-principal and quantitative authority.} The current IR is a
    one-principal, one-action Boolean instance. A general treatment needs
    principal sets, action families, quantitative permissions, delegation
    chains, revocation, and a cost model for provenance precision.
  \item \emph{Pure versus mixed expressiveness.} The decomposition in
    Section~\ref{sec:baseline} is one direction only. We do not know whether
    every mixed defence can be represented by a pure system-level rule and a
    pure model-level rule with the same security and utility. That equivalence
    is an open question, not a claim of this paper.
  \item \emph{Multi-agent trust.} Extending the principal model to agents that
    delegate to, observe, or collude with other agents raises trust
    transitivity, confused-deputy, cross-agent revocation, and inter-agent
    provenance questions. The formal result is single-agent in this sense: several principals may
    influence one action, but the model does not contain agents that delegate,
    observe, or collude with other agents. The current theorem treats each agent
    as a principal and does not establish a multi-agent trust result.
  \item \emph{Explanations and human factors.} Deployments need usable
    explanations of why an action was denied, how a principal entered the
    context, and what a user can do to grant or revoke authority.
  \item \emph{Production isolation and attestation.} Provider adapters are not
    isolation boundaries. Binding the mediator to an attested execution
    environment, making the certificate channel tamper-evident, and closing the
    recheck-to-effect race are open engineering tasks.
\end{itemize}

\section{Limitations}
\label{sec:limitations}

\paragraph{Side channels and timing.}
We do not cover side-channel or timing attacks. The authority rule constrains
which action is admitted; it does not bound what a principal can infer from
latency, output order, error messages, or resource use. Authorised-read safety
is not noninterference, and the finite IR does not model covert channels. A
per-principal integrity result would also need to show that one principal's
influence cannot alter another principal's authorised actions except through
recorded provenance; the PE property is action-level and does not establish that
stronger directional condition.

\paragraph{Trusted computing base.}
The result assumes sound authentication, complete mediation, accurate
provenance, current policy state, action binding, and a known non-empty decision
context. A2 requires the provenance store to record every input that causally
influences the model's proposal or the action, including control inputs that
select the tool, branch, recipient, or plan step, and any relay or message filter
that controls what reaches the model. Identity is assumed, not established: a
relay that forwards a principal's data must be recorded as an author.

A concrete failure is an attacker-controlled gateway that forwards a benign
message from a trusted principal and suppresses or reorders a conflicting
message from another principal. Provenance records the delivered author, not the
gateway. The gateway's selection changes the model's output, so the recorded set
under-approximates causal influence and A2 fails. The theorem does not apply to
that execution. A compromised executor, provenance adapter, or policy port is
similarly outside the model. Certificates detect stale decisions and mutated
actions, but they are not a substitute for an attested execution environment.

\paragraph{Policy adequacy.}
The theorem does not establish that the ACS encodes the organisation's intended
restrictions. A policy that is too permissive yields a permitted escalation; a
policy that is too restrictive yields false denials. Both are assurance
questions outside the confinement proof.

\paragraph{Utility and liveness.}
The result is safety-only. Conservative provenance and independent policy
dimensions can deny admissible tasks. The paper does not claim that a planner
finds a plan, that a model completes a task, or that a deployment preserves
utility. Exposure planning is a bounded planning model, not an operational
planner guarantee.

\paragraph{Delegation and messages.}
Delegation is specified and verified over finite models but is not enabled at
runtime. Messages are disclosure events governed by visibility and consent, not
by the per-principal authorisation relation; they lie outside the confinement
theorem.

\paragraph{Evidence scope.}
The AgentDojo runs are read-only and partial; the SLED and SLED-V results are
bounded statements about declared models and fixtures. They do not establish
implementation conformance, unbounded security, or superiority over a deployed
defence.

\section{Conclusion}
\label{sec:conclusion}

The contribution is a principal-sensitive enforcement design for privilege
escalation in LLM agents. We formulate PE as the system-level objective, define
ITES as an authority rule over the provenance-recorded influencing principals,
and prove conditional confinement under A1--A5. ITES takes authority from the
organisation's existing ACS and keeps model judgement outside the authority
trusted computing base. The result is conditional on accurate provenance, stated as A2.

Conflux extends the design toward more expressive authority. It separates read,
authorisation, visibility, and consent, and adds argument-aware policy,
disclosure, attribution, delegation, restriction, and exposure planning. SLED
evaluates a declared finite model of a defence; SLED-V checks a serialisable IR
with bounded solver routes. Prior systems use related ingredients: provenance tracking, information-flow
labels, and effect-time policy checks. They do not define authority as the
intersection of the current ACS permissions of every provenance-recorded
influencing principal. That principal-set formulation and the confinement result
are the contribution here.

Finite-model results are statements about declared models, and the AgentDojo runs
exercise read-only paths. The paper's claim is the conditional confinement result
and the evaluation instruments that make it checkable.

\bibliographystyle{plainnat}
\bibliography{references}

\appendix

\section{SLED results}
\label{app:sled}

The archived evidence below is from SLED, the black-box evaluator, unless a
subsection says otherwise. Each SLED result is a bounded statement about a
declared finite model. We report the three newest archived SLED runs: the
future-code ablation, the final-depth batch sweep, and the environment-complexity
series. A provenance-precision fixture follows separately; it is a symbolic
measurement over delegation fixtures, not a SLED verdict run. Older archived
trace counts used a different enumeration and are deliberately omitted. In the
tables, \emph{safe} means the finite state space was exhausted with no violation
and \texttt{truncated=False}; it does not mean an unbounded proof.

\begin{figure}[t]
\centering
\begin{tikzpicture}
\begin{axis}[
  ybar,
  width=0.96\linewidth,
  height=5.4cm,
  ymin=0, ymax=112,
  ylabel={task instances (\%)},
  symbolic x coords={sequential safe,resolved safe,assumed violation,actual violation,benign goal reached,task unresolved},
  xtick=data,
  x tick label style={rotate=18,anchor=east,font=\small},
  legend style={at={(0.5,-0.34)},anchor=north,legend columns=2,font=\small,draw=none},
  bar width=10pt,
  enlarge x limits=0.12,
  axis line style={draw=black!55},
  tick style={draw=black!55},
]
\addplot[fill=blue!55,draw=blue!70] coordinates {
  (sequential safe,40)
  (resolved safe,0)
  (assumed violation,60)
  (actual violation,0)
  (benign goal reached,40)
  (task unresolved,60)
};
\addplot[fill=orange!65,draw=orange!80] coordinates {
  (sequential safe,40)
  (resolved safe,35)
  (assumed violation,0)
  (actual violation,25)
  (benign goal reached,100)
  (task unresolved,0)
};
\legend{future code omitted,future code modeled}
\end{axis}
\end{tikzpicture}
\caption{Controlled SLED future-code ablation over 20 task instances: eight
sequential, seven future-dependent safe, and five future-dependent unsafe.
Without expansion the evaluator conservatively assumes interference on the
twelve future-dependent tasks; the eight sequential tasks remain safe and
complete. With expansion it resolves those tasks and reaches the benign goal on
all twenty; five of those also expose an unsafe branch, so safe completion
without a violation is fifteen of twenty. The fixture is a check on the
evaluation obligation, not an empirical estimate of a deployed defence. The
plotted values are transcribed from the archived ablation run records.}
\label{fig:future-code-ablation}
\end{figure}
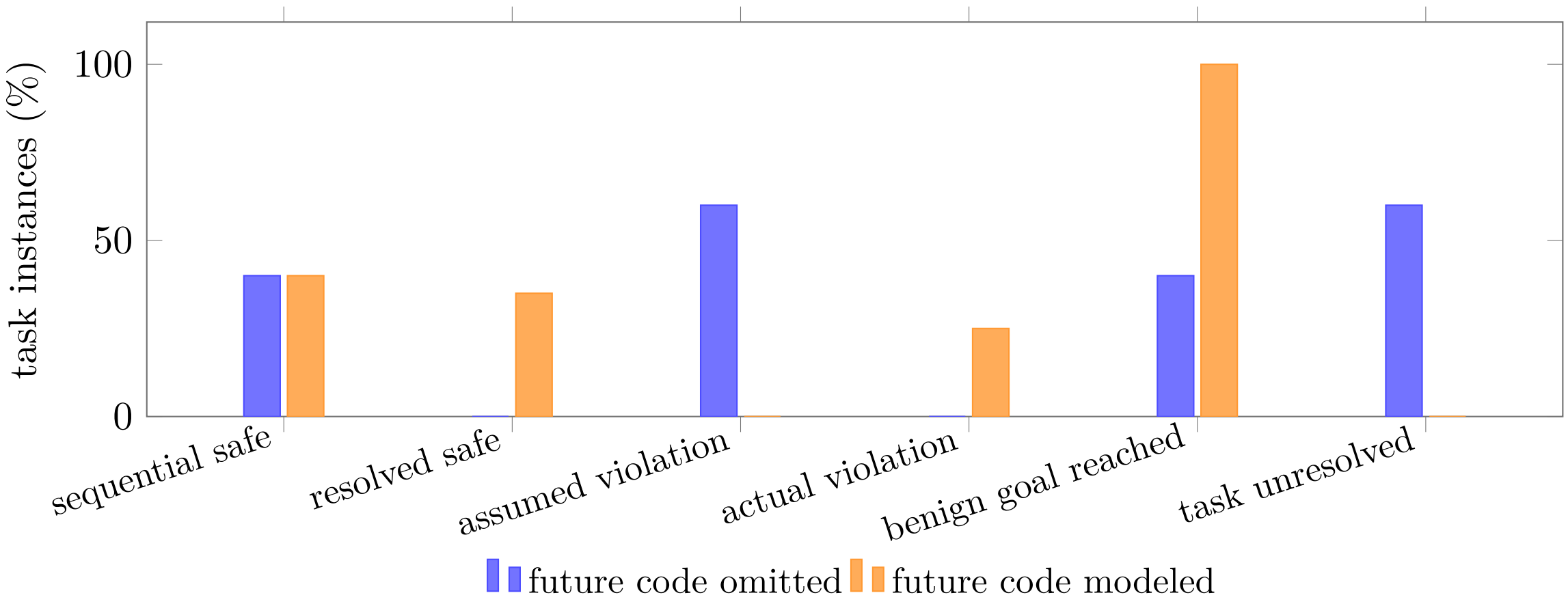

\paragraph{Future-code ablation.}
The evaluator must expand the code that a nested execution would run.  The
controlled fixture in Figure~\ref{fig:future-code-ablation} separates eight
sequential tasks, which need no nested-code model, from twelve
future-dependent tasks, which do.  With expansion disabled, the fixture
conservatively assumes interference on every future-dependent task: the
missing code could contain the unsafe action, so the evaluator reports a
potential violation and does not claim task completion.  The eight sequential
tasks remain safe and complete.  With expansion enabled, the evaluator
resolves the twelve future-dependent tasks: seven are safe and complete, and
five expose an unsafe branch admitted by the fixture's deliberately buggy
nested defence.  The figure is an illustration of the evaluation obligation,
not a measurement of a deployed model; its point is that a clean security
count without future code would be unsupported, and that expansion is needed
to distinguish safe from unsafe future-dependent tasks.

\paragraph{Final-depth batch bound.}
The future-code ablation above isolates the expansion step. Table~\ref{tab:sled-final-batch-sweep} isolates the final batch grammar. The archived prototype caps the final model call at three primitive actions; we relaxed that cap in the current evaluator from three to six on five small configurations. Every run is \texttt{safe} and reports \texttt{truncated=False}: the finite state
space was exhausted with no violation. State counts do not change, while transitions rise with the number of admitted final batches: on \texttt{p2-a6-d3} and \texttt{p3-a6-d3}, transitions rise from 2,024 at bound three to 2,696 at bound six. The check is bounded and uses the archived configurations, so it does not establish an unbounded result; it shows that the three-action cap is not what makes these bounded runs pass.

\begin{table}[t]
\centering
\small
\begin{tabular}{lrrrrrl}
\toprule
\textbf{Config} & \textbf{States} & \textbf{B3} & \textbf{B4} & \textbf{B5} & \textbf{B6} & \textbf{Verdict} \\
\midrule
\texttt{p2-a3-d3} & 30 & 884 & 884 & 884 & 884 & safe \\
\texttt{p2-a4-d4} & 58 & 5{,}252 & 5{,}308 & 5{,}308 & 5{,}308 & safe \\
\texttt{p2-a6-d3} & 30 & 2{,}024 & 2{,}444 & 2{,}654 & 2{,}696 & safe \\
\texttt{p3-a4-d4} & 58 & 5{,}252 & 5{,}308 & 5{,}308 & 5{,}308 & safe \\
\texttt{p3-a6-d3} & 30 & 2{,}024 & 2{,}444 & 2{,}654 & 2{,}696 & safe \\
\bottomrule
\end{tabular}
\caption{Final-depth batch-bound sweep on five archived configurations. The final model call may propose up to the bound shown. All runs report \texttt{truncated=False} and verdict \texttt{safe}. State counts are unchanged by the bound. Counts are read from the archived run records for this sweep. This is a bounded check on the tested finite models, not an unbounded proof.}
\label{tab:sled-final-batch-sweep}
\end{table}

\paragraph{Environment complexity.}
Table~\ref{tab:sled-complexity} records the unique principal, input, and action
counts for each generated environment and each attempted depth. It also records
the raw complete-trace count at the cutoff. The counts are not a simple product
of the three structural dimensions: authorship, permissions, and the option
grammar determine which proposals are reachable. Rows marked \emph{cutoff}
stopped at the per-stage time limit after a whole number of traces, so they are
lower bounds: depth-3 stages used a 3{,}600\,s limit, while depth-4 stages used
the remaining stage budget. Depth-3 and depth-4 runs are separate measurements
because the depth bound differs; the option sets are the same at both depths,
so a depth-3 trace is not a prefix of a depth-4 trace. E6 and E7 have the same depth-3 count despite different principal
counts; the table reports that coincidence rather than treating the labels as
evidence of behavioural diversity.

\begin{table}[t]
\centering
\footnotesize
\resizebox{\linewidth}{!}{%
\begin{tabular}{llrrrrrl}
\toprule
\textbf{ID} & \textbf{Environment} & \textbf{U} & \textbf{I} & \textbf{A} & \textbf{Depth} & \textbf{Traces} & \textbf{Status} \\
\midrule
E1 & \texttt{p10-a10-d8} & 10 & 8 & 10 & 3 & 188{,}834{,}048 & cutoff \\
E2 & \texttt{p10-a8-d5} & 10 & 5 & 8 & 3 & 53{,}089{,}273 & complete \\
E2 & \texttt{p10-a8-d5} & 10 & 5 & 8 & 4 & 170{,}196{,}736 & cutoff \\
E3 & \texttt{p12-a10-d6} & 12 & 6 & 10 & 3 & 224{,}106{,}240 & cutoff \\
E4 & \texttt{p16-a8-d7} & 16 & 7 & 8 & 3 & 233{,}644{,}544 & cutoff \\
E5 & \texttt{p20-a10-d7} & 20 & 7 & 10 & 3 & 212{,}929{,}024 & cutoff \\
E6 & \texttt{p20-a4-d5} & 20 & 5 & 4 & 3 & 5{,}844{,}441 & complete \\
E6 & \texttt{p20-a4-d5} & 20 & 5 & 4 & 4 & 158{,}657{,}280 & cutoff \\
E7 & \texttt{p24-a4-d5} & 24 & 5 & 4 & 3 & 5{,}844{,}441 & complete \\
E7 & \texttt{p24-a4-d5} & 24 & 5 & 4 & 4 & 161{,}091{,}840 & cutoff \\
E8 & \texttt{p3-a3-d9} & 3 & 9 & 3 & 3 & 239{,}979{,}008 & cutoff \\
E9 & \texttt{p4-a4-d8} & 4 & 8 & 4 & 3 & 169{,}961{,}728 & cutoff \\
E10 & \texttt{p6-a5-d8} & 6 & 8 & 5 & 3 & 202{,}124{,}544 & cutoff \\
E11 & \texttt{p6-a6-d9} & 6 & 9 & 6 & 3 & 258{,}883{,}584 & cutoff \\
E12 & \texttt{p8-a6-d8} & 8 & 8 & 6 & 3 & 202{,}214{,}912 & cutoff \\
\bottomrule
\end{tabular}}
\caption{SLED environment complexity and raw trace counts. $U$ is the number of unique principals, $I$ the number of unique inputs or data items, and $A$ the number of unique actions. All runs used \texttt{all\_readers=True} and one initial data item. Rows marked \emph{cutoff} are whole complete traces at the per-stage time limit---3{,}600\,s for depth 3 and the remaining stage budget for depth 4---hence lower bounds, not estimates. Depth-3 and depth-4 counts are independent because the depth bound differs; the option sets are the same at both depths, so traces are not nested across depths. E6 and E7 illustrate why labels alone are not a diversity guarantee: their depth-3 counts coincide. Counts are read from the archived per-environment stage records.}
\label{tab:sled-complexity}
\end{table}

\begin{figure}[t]
\centering
\begin{tikzpicture}
\begin{axis}[
  ybar,
  width=0.98\linewidth,
  height=5.4cm,
  ymin=0, ymax=8.9,
  ylabel={$\log_{10}$(complete traces)},
  xtick={1,2,3,4,5,6,7,8,9,10,11,12},
  xticklabels={E1,E2,E3,E4,E5,E6,E7,E8,E9,E10,E11,E12},
  x tick label style={font=\scriptsize},
  y tick label style={font=\scriptsize},
  legend style={at={(0.5,-0.28)},anchor=north,legend columns=3,font=\scriptsize,draw=none},
  bar width=4.2pt,
  enlarge x limits=0.03,
  axis line style={draw=black!55},
  tick style={draw=black!55},
]
\addplot[fill=blue!55,draw=blue!70] coordinates {
  (1,0) (2,7.725) (3,0) (4,0) (5,0) (6,6.767) (7,6.767)
  (8,0) (9,0) (10,0) (11,0) (12,0)
};
\addplot[fill=blue!15,draw=blue!55] coordinates {
  (1,8.276) (2,0) (3,8.350) (4,8.369) (5,8.328) (6,0) (7,0)
  (8,8.380) (9,8.230) (10,8.306) (11,8.413) (12,8.306)
};
\addplot[fill=orange!35,draw=orange!70] coordinates {
  (1,0) (2,8.231) (3,0) (4,0) (5,0) (6,8.200) (7,8.207)
  (8,0) (9,0) (10,0) (11,0) (12,0)
};
\legend{depth 3 complete,depth 3 cutoff,depth 4 cutoff}
\end{axis}
\end{tikzpicture}
\caption{State growth across the 12 generated environments. The vertical axis is
the base-10 logarithm of the raw complete-trace count. Light bars are lower
bounds from the per-stage cutoff. Only three environments completed depth 3;
the depth-4 bars are cutoff counts for those same environments. Counts are read
from the archived per-environment stage records. The figure
shows that raw counts vary by orders of magnitude across structural
configurations; it is not a fitted complexity law.}
\label{fig:sled-complexity}
\end{figure}
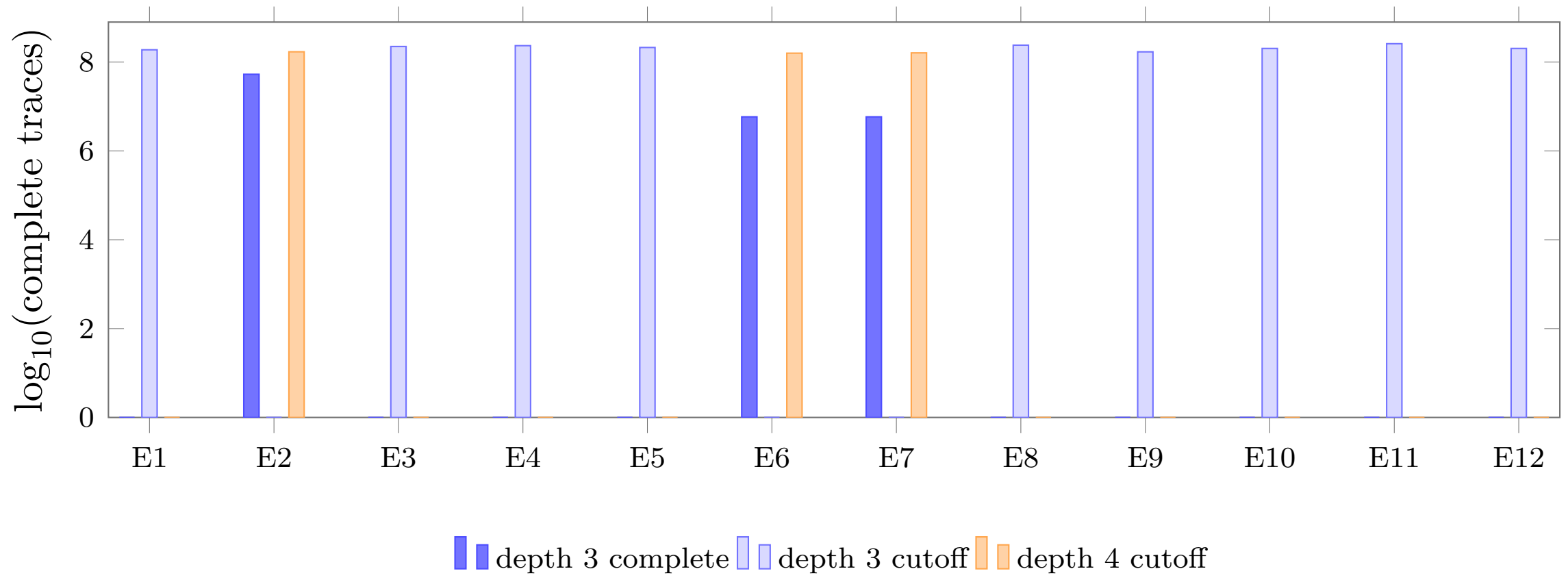

\subsection{Provenance precision (separate from SLED)}
\label{app:provenance}

This fixture measures a symbolic provenance rule, not a SLED state-space
verdict. It is included because the rule is used by the defence, but its counts
are not a SLED outcome.

\begin{table}[t]
\centering
\small
\resizebox{\linewidth}{!}{\begin{tabular}{llrrrr}
\toprule
\textbf{Provenance rule} & \textbf{Provenance set} & \textbf{Mean size} & \textbf{Secure tasks completed} & \textbf{False denials} & \textbf{Vulnerable executions} \\
\midrule
C0 requester only (negative control) & requester & 1.00 & 208 of 208 & 0 of 208 & 48 of 48 \\
C1 may-have-written provenance & possible authors & 8.62 & 173 of 208 & 35 of 208 & 0 of 48 \\
C2 authenticated provenance & actual contributors & 3.08 & 208 of 208 & 0 of 208 & 0 of 48 \\
\bottomrule
\end{tabular}}
\caption{Capability retention and over-blocking for three provenance rules over 351 symbolic delegation fixtures (208 securely achievable, 95 securely blocked, 48 vulnerable), three repetitions per parameter cell. C0 is the negative control: it authorises the requester alone, so its row shows the vulnerability cost of omitting confinement, not the cost of enforcing it. From the conservative rule to authenticated provenance, denials of securely achievable work fall from 35 to 0, the mean size of the computed provenance set falls from 8.62 to 3.08, and 0 of 351 authenticated rows both authorise and violate the confinement rule. The measurement is one-sided: the rows count the denials the rule produces, and the bundle holds no refusal counter and no model, so a refusal an agent would add on top is uncounted. Counts are read from the archived provenance-precision fixture records.}
\label{tab:provenance-precision}
\end{table}

\section{SLED-V and IR results}
\label{app:sledv}

The semantic-conformance table checks that the reference interpreter and two
optional solver backends decide the same bounded IR variants, including a
negative control. It shows that the encodings agree on the tested variants; it
does not show sensitivity to defect interactions.

\begin{table}[t]
\centering
\small
\begin{tabular}{lr}
\toprule
\textbf{Quantity} & \textbf{Count} \\
\midrule
IR variants checked & 11 \\
Decided by solver A & 11 \\
Decided by solver B & 11 \\
Refuted by the reference interpreter & 10 \\
Reference and solver A agree & 11 \\
Reference and solver B agree & 11 \\
All three agree & 11 \\
Negative control & holds \\
\bottomrule
\end{tabular}
\caption{Semantic conformance over the same bounded IR variants: the reference interpreter, two optional solver backends decide the same variants and agree on every decision, refuted variants included, and the negative control confirms the harness can still fail. Counts are read from the archived result file, so the table cannot drift from the evidence it cites.}
\label{tab:semantic-conformance}
\end{table}

\paragraph{Scope of the conformance check.}
The check tests the encoding, not sensitivity to defect interactions, and it is
not a proof of the delegation semantics.

\end{document}